\documentstyle[twocolumn,aps]{revtex}

\begin{document}
\draft
\title{Reply on ``Fluctuation-dissipation considerations for phenomenological
damping models for ferromagnetic thin films'' [N. Smith, J. Appl. Phys. {\bf %
92}, 3877 (2002)]}
\author{Vladimir L. Safonov and H. Neal Bertram}
\address{Center for Magnetic Recording Research, \\
University of California - San Diego, \\
9500 Gilman Drive, La Jolla, CA 92093-0401}
\date{\today}
\maketitle

\begin{abstract}
We show that the critique of our recent papers presented in the
abovementioned paper (NS) appeals to an incorrect mathematical analogy
between electrical circuits and linear magnetization dynamics, improperly
uses classical concepts of normal modes and basic equations, gives
inconsistent results and therefore comes to incorrect conclusions.
\end{abstract}

\bigskip 

The study of linear stochastic magnetization dynamics is of great importance
in applications to nano-magnetic devices based on ultra-thin films.
Gyromagnetic magnetization motion around an effective field is randomly
forced by fluctuations on spins by means of interaction with a thermal bath
(phonons, magnons, conduction electrons, impurities, etc.). The role of the
thermal bath can be approximately reduced to terms describing relaxation and
random fields in the magnetization dynamic equations.

For simplicity, the conventional theoretical approach is purely
phenomenological. It is based on the Landau-Lifshitz equation \cite{landau}
with random fields, or its modification in a Gilbert form \cite{gilbert}.
This approach has the defect that the phenomenological damping term in
Landau-Lifshitz (and Gilbert) equation was introduced just for the case of
high magnetic symmetry (axial symmetry). An applicability of an isotropic
damping together with corresponding noise to strongly anisotropic system
such as thin film has not been proven \cite{comment0}.

In our recent papers \cite{bjs},\cite{SafBertNoise} we have developed a
theoretical approach based upon the representation of the magnetization
dynamics as the motion of damped nonlinear oscillator driven by a random
force (thermal fluctuations). The oscillator model is a convenient tool \cite
{sparks} to establish a ``bridge'' between the microscopic physics, where
the oscillator variables $a^{\ast }$ and $a$ naturally describe spin
excitations (as creation and annihilation operators), and the macroscopic
stochastic magnetization dynamics for normal modes. We have calculated the
magnetization noise spectrum in a thin film for a physical loss mechanisms 
\cite{mechanisms} and shown that our result differs from that of obtained by
the phenomenological approach and is in agreement with experiment \cite
{experiment}.

Recently Neil Smith (the paper cited in the title, later, NS) criticized our
approach claiming that ``a proper understanding of the FDT
(fluctuation-dissipation theorem) ... may have been complicated by several
recent papers \cite{bjs},\cite{SafBertNoise}''. In this Reply we argue that
this critique 1) appeals to an incorrect mathematical analogy between
electrical circuits and linear magnetization dynamics, 2) improperly uses
classical concepts of normal modes and basic equations claiming simultaneous
diagonalization of three hermitian matrices, and 3) gives inconsistent
results. 

\bigskip 

1. {\it Incorrect mathematical analogy. }Any analogy in physics is based on
a similar structure of mathematical equations. The basic equations for
linear electric circuits have the form (here and later we use notations of
NS):

\begin{equation}
\sum_{j=1}^{N}\left[ M_{ij}\frac{d^{2}X_{j}}{dt^{2}}+D_{ij}\frac{dX_{j}}{dt}%
+K_{ij}X_{j}\right] =F_{i}(t).  \label{Electrical}
\end{equation}
This is a set of the second order differential equations, where the scalar
variables  $X_{j}$ describe electric charges, $F_{i}(t)$ are external
voltages and symmetric matrices describe inductances ($M_{ij}$), resistors ($%
D_{ij}$) and capacitors ($K_{ij}$).

On the other hand, linear magnetization dynamics do not contain inertial
terms ($M_{ij}=0$) and do contain an antisymmetric gyromagnetic matrix $%
G_{ij}$:

\begin{equation}
\sum_{j=1}^{N}\left[ (D_{ij}-G_{ij})\frac{dX_{j}}{dt}+K_{ij}X_{j}\right]
=F_{i}(t).  \label{Magnetic}
\end{equation}
This is a set of first order differential equations and there is no
mathematical analog for $G_{ij}$ in a linear electric scheme. In order to
describe $N$ coupled magnetic oscillators (spin waves) each dynamic variable 
$X_{j}$ should be chosen as a column (not a scalar), containing two
transverse magnetization components in the $j$-th micromagnetic cell:

\begin{equation}
X_{j}=%
{m_{x,j} \choose m_{y,j}}%
.  \label{mvariable}
\end{equation}

In this formulation there is no direct mathematical analogy between linear
electrical circuits (\ref{Electrical}) and linear magnetization dynamics (%
\ref{Magnetic}). Thus, without loss of generality, the compilation of
``electrical analogies'' (Sec.III in NS) has no mathematical consequences
for magnetic systems \cite{comment1}.

\bigskip 

2. {\it Incorrect claim of simultaneous diagonalization of three hermitian
matrices (including damping }$\stackrel{\longleftrightarrow }{{\bf D}}$){\it %
.} We think that this claim is just a simple misunderstanding in NS of the
normal mode (spin wave) concept in magnetic dynamics. Spin waves are a
principal concept of a conservative magnetic system where the energy of the
linear magnetic oscillations can be diagonalized and represented as a set of
independent harmonic oscillators (see, e.g., \cite{sparks}). Uniform
rotation of the magnetization corresponds to a spin wave with zero momentum.
Spin waves interact with each other and with other degrees of freedom
(elastic waves, conduction electrons, etc.). Spin wave damping appears as a
result of reducing (averaging) these interactions (interaction with a
thermal bath). Therefore the reduced equations can not have a general
structure. The damping matrix is specific to the magnetic system and can not
have an arbitrary form, as assumed in NS. 

\bigskip 

3. {\it Inconsistency of obtained ``general'' results. }Let us check the
general results obtained in NS for the fluctuation-dissipation theorem as
used in Gilbert and Bloch-Bloembergen dynamics. In NS the Gilbert equation
is written in the form:

\begin{equation}
(\stackrel{\longleftrightarrow }{{\bf D}}-\stackrel{\longleftrightarrow }{%
{\bf G}})\cdot \frac{d\overrightarrow{{\bf m}}}{dt}+\stackrel{%
\longleftrightarrow }{{\bf H}}\cdot \overrightarrow{{\bf m}}=\overrightarrow{%
{\bf h}}(t),  \label{Gilbert}
\end{equation}
where 
\begin{equation}
\stackrel{\longleftrightarrow }{{\bf D}}=\frac{\alpha }{\gamma }\left( 
\begin{array}{cc}
1 & 0 \\ 
0 & 1
\end{array}
\right) ,\quad \stackrel{\longleftrightarrow }{{\bf G}}=\frac{1}{\gamma }%
\left( 
\begin{array}{cc}
0 & 1 \\ 
-1 & 0
\end{array}
\right)   \label{DG}
\end{equation}
and $\stackrel{\longleftrightarrow }{{\bf H}}$ describes effective magnetic
fields. The corresponding thermal fluctuation fields are given by: 
\begin{equation}
\langle \overrightarrow{{\bf h}}(t_{0}+\tau )\overrightarrow{{\bf h}}%
(t_{0})\rangle _{ij}=\frac{2kT\alpha }{\gamma M_{s}\Delta V}\left( 
\begin{array}{cc}
1 & 0 \\ 
0 & 1
\end{array}
\right) \delta _{ij}\delta (\tau ),  \label{GilbertFDT}
\end{equation}
referred as FDT in NS. It is well known that the Gilbert equation can be
rewritten in a mathematically equivalent form of the Landau-Lifshitz
equation. This procedure is most simple in the case of small damping when we
can neglect the higher order damping terms (such as $\alpha ^{2}$). We can
find the magnetization derivative on time from Eq.(\ref{Gilbert}) without
damping

\begin{equation}
\frac{d\overrightarrow{{\bf m}}}{dt}=\left( -\stackrel{\longleftrightarrow }{%
{\bf G}}\right) ^{-1}\cdot \left( -\stackrel{\longleftrightarrow }{{\bf H}}%
\cdot \overrightarrow{{\bf m}}+\overrightarrow{{\bf h}}(t)\right)
\label{Ggyrom}
\end{equation}
and represent the damping term as

\begin{eqnarray}
\stackrel{\longleftrightarrow }{{\bf D}}\cdot \frac{d\overrightarrow{{\bf m}}%
}{dt} &=&\stackrel{\longleftrightarrow }{{\bf D}}\cdot \left( -\stackrel{%
\longleftrightarrow }{{\bf G}}\right) ^{-1}\cdot \left( -\stackrel{%
\longleftrightarrow }{{\bf H}}\cdot \overrightarrow{{\bf m}}+\overrightarrow{%
{\bf h}}(t)\right)   \nonumber \\
&=&\alpha \gamma \left( -\stackrel{\longleftrightarrow }{{\bf G}}\cdot 
\stackrel{\longleftrightarrow }{{\bf H}}\cdot \overrightarrow{{\bf m}}+%
\stackrel{\longleftrightarrow }{{\bf G}}\cdot \overrightarrow{{\bf h}}%
(t)\right) .  \label{da}
\end{eqnarray}
Here we utilize (\ref{DG}) and\ take into account that $\left( -\stackrel{%
\longleftrightarrow }{{\bf G}}\right) ^{-1}=\gamma ^{2}\stackrel{%
\longleftrightarrow }{{\bf G}}$. Neglecting small $\alpha \gamma \stackrel{%
\longleftrightarrow }{{\bf G}}\cdot \overrightarrow{{\bf h}}(t)$, the Eq.(%
\ref{Gilbert}) may be written as:

\begin{equation}
-\stackrel{\longleftrightarrow }{{\bf G}}\cdot \frac{d\overrightarrow{{\bf m}%
}}{dt}+\left( \stackrel{\longleftrightarrow }{{\bf H}}-\stackrel{%
\longleftrightarrow }{{\bf G}}\cdot \alpha \gamma \stackrel{%
\longleftrightarrow }{{\bf H}}\right) \cdot \overrightarrow{{\bf m}}=%
\overrightarrow{{\bf h}}(t).  \label{Gilbsmall}
\end{equation}

On the other hand, in NS the Bloch-Bloembergen equation has been written as:

\begin{equation}
-\stackrel{\longleftrightarrow }{{\bf G}}\cdot \frac{d\overrightarrow{{\bf m}%
}}{dt}+(\stackrel{\longleftrightarrow }{{\bf H}}-\stackrel{%
\longleftrightarrow }{{\bf G}}/T_{2})\cdot \overrightarrow{{\bf m}}=%
\overrightarrow{{\bf h}}(t)  \label{BB}
\end{equation}
with corresponding FDT of the form:

\begin{equation}
\langle \overrightarrow{{\bf h}}(t_{0}+\tau )\overrightarrow{{\bf h}}%
(t_{0})\rangle =\frac{kT}{\gamma M_{s}\Delta V}\left( 
\begin{array}{cc}
0 & -1 \\ 
1 & 0
\end{array}
\right) \delta _{ij}\frac{{\rm sgn}(\tau )}{T_{2}}.  \label{FDTBB}
\end{equation}

We see that Eqs.(\ref{Gilbsmall}) and (\ref{BB}) have similar forms. They
coincide, for example, in the case

\begin{equation}
\stackrel{\longleftrightarrow }{{\bf H}}=H_{0}\left( 
\begin{array}{cc}
1 & 0 \\ 
0 & 1
\end{array}
\right)   \label{isotropicH}
\end{equation}
and $1/T_{2}=\alpha \gamma H_{0}$. In this case their FDT relations (\ref
{GilbertFDT}) and (\ref{FDTBB}) are distinctly different from each other.
This fact simply indicates that NS results are inconsistent.

\bigskip 

This work was partly supported by matching funds from the Center for
Magnetic Recording Research at the University of California - San Diego and
CMRR incorporated sponsor accounts.


\begin{references}
\bibitem{comment1}  In principle, it would be interesting to construct a
correct mathematical analogy between linear electrical circuits and linear
magnetization dynamics using appropriate choice of variables in magnetic
system. This, however, has not been done in NS.
\end{references}
\end{document}